\documentclass[aps,twocolumn,superscriptaddress,pra]{revtex4-1}

\usepackage{amsmath,amssymb,graphicx,color,verbatim,soul,braket,upgreek}
\usepackage{etoolbox}
\usepackage{hyperref}

\begin{document}

\renewcommand{\figurename}{\textbf{Figure}}
\title{Observation of universal dynamics in a spinor Bose gas far from equilibrium}

\author{Maximilian Pr\"ufer}
\affiliation{Kirchhoff-Institut f\"ur Physik, Universit\"at Heidelberg, Im Neuenheimer Feld 227, 69120 Heidelberg, Germany}
\author{Philipp Kunkel}
\affiliation{Kirchhoff-Institut f\"ur Physik, Universit\"at Heidelberg, Im Neuenheimer Feld 227, 69120 Heidelberg, Germany}
\author{Helmut Strobel}
\affiliation{Kirchhoff-Institut f\"ur Physik, Universit\"at Heidelberg, Im Neuenheimer Feld 227, 69120 Heidelberg, Germany}
\author{Stefan Lannig}
\affiliation{Kirchhoff-Institut f\"ur Physik, Universit\"at Heidelberg, Im Neuenheimer Feld 227, 69120 Heidelberg, Germany}
\author{Daniel Linnemann}
\affiliation{Kirchhoff-Institut f\"ur Physik, Universit\"at Heidelberg, Im Neuenheimer Feld 227, 69120 Heidelberg, Germany}
\author{Christian-Marcel Schmied}
\affiliation{Kirchhoff-Institut f\"ur Physik, Universit\"at Heidelberg, Im Neuenheimer Feld 227, 69120 Heidelberg, Germany}
\author{J\"urgen Berges}
\affiliation{Institut f\"ur Theoretische Physik, Universit\"at Heidelberg, Philosophenweg 16, 69120 Heidelberg, Germany}
\author{Thomas Gasenzer}
\affiliation{Kirchhoff-Institut f\"ur Physik, Universit\"at Heidelberg, Im Neuenheimer Feld 227, 69120 Heidelberg, Germany}
\author{Markus K.\ Oberthaler}
\affiliation{Kirchhoff-Institut f\"ur Physik, Universit\"at Heidelberg, Im Neuenheimer Feld 227, 69120 Heidelberg, Germany}

\date{\today}

\maketitle
\textbf{
The dynamics of quantum systems far from equilibrium represents one of the most challenging problems in theoretical many-body physics \citep{Polkovnikov2011,Eisert2018}. 
While the evolution is in general intractable in all its details, relevant observables can become insensitive to microscopic system parameters and initial conditions. 
This is the basis of the phenomenon of universality. 
Far from equilibrium, universality is identified through the scaling of the spatio-temporal evolution of the system, captured by universal exponents and functions. 
Theoretically, this has been studied in examples as different as the reheating process in inflationary universe cosmology \cite{Kofman1994,Micha2003}, the dynamics of nuclear collision experiments described by quantum chromodynamics \cite{Baier2001,Berges2014}, or the post-quench dynamics in dilute quantum gases in non-relativistic quantum field theory \cite{Lamacraft2007,Barnett2011,Hofmann2014,Orioli2015,Williamson2016}. 
However, an experimental demonstration of such scaling evolution in space and time in a quantum many-body system is lacking so far. 
Here we observe the emergence of universal dynamics by evaluating spatially resolved spin correlations in a quasi one-dimensional spinor Bose-Einstein condensate \cite{Sadler2006,Kronjager2010,Bookjans2011,De2014,Nicklas2015}. 
For long evolution times we extract the scaling properties from the spatial correlations of the spin excitations.
From this we find the dynamics to be governed by transport of an emergent conserved quantity towards low momentum scales. 
Our results establish an important class of non-stationary systems whose dynamics is encoded in time-independent scaling exponents and functions signaling the existence of non-thermal fixed points \cite{Berges2008,Nowak2010,Orioli2015}. 
We confirm that the non-thermal scaling phenomenon involves no fine-tuning, by preparing different initial conditions and observing the same scaling behaviour. 
Our analog quantum simulation approach provides the basis to reveal the underlying mechanisms and characteristics of non-thermal universality classes. 
One may use this universality to learn, from experiments with ultra-cold gases, about fundamental aspects of dynamics studied in cosmology and quantum chromodynamics.
}


Isolated quantum many-body systems offer particularly clean settings for studying fundamental properties of the underlying unitary time evolution~\cite{Bloch2012}. 
For systems initialised far from equilibrium different scenarios have been identified, including the occurence of many-body oscillations \cite{Hung2013} and revivals~\cite{Rauer2018}, the manifestation of many-body localisation~\cite{Schreiber2015}, and quasi-stationary behaviour in a prethermalised stage of the evolution~\cite{Gring2012}.

Here we observe a new scenario associated to the notion of non-thermal fixed points.
This is illustrated schematically in Fig.\,1a: 
Starting from a class of far-from-equilibrium initial conditions, the system develops a universal scaling behaviour in time and space.
This is a consequence of the effective loss of details about initial conditions and system parameters long before a quasi-stationary or equilibrium situation may be reached.
The transient scaling behaviour is found to be governed by the transport of an emergent collective conserved quantity towards low momentum scales.

For our experimental study we employ an elongated Bose-Einstein condensate of $\sim70,\!000$ $^{87}$Rb atoms.
We use the $F=1$ hyperfine manifold with its three magnetic sublevels $m_\text{F} = 0,\pm1$ as a spin-1 system with ferromagnetic interactions \citep{StamperKurn2013}.
Initially, all atoms are prepared in the $m_\text{F} = 0$ sublevel, forming a spinor condensate with zero spin length.
The dynamics is initiated by instantaneously changing the energy splitting of the $F = 1$ magnetic sublevels by means of microwave dressing (see Methods). 
Consequently spin excitations develop in the $F_{x}$--$F_{y}$-plane \citep{Sadler2006} as sketched in Fig.\,1b.
Our experimental setup allows the extraction of the spin distribution in terms of the spin component $\hat{F}_x(y) =  \, ( \hat{\psi}_0(y)\big[\hat{\psi}^\dag_{+1}(y)+\hat{\psi}^\dag_{-1}(y)\big] + \text{h.c.} )  \, /\sqrt{2}$, where $\hat{\psi}_m^\dag(y)$ is the creation operator of an atom in the magnetic sublevel $m$ at position $y$.
At a given time $t$ this is achieved by a spin rotation from the $F_{x}$--$F_{y}$-plane to the $F_{z}$-direction and subsequently detecting the atomic density difference $F_{z}(y)=n_{+1}(y)-n_{-1}(y)$ (see Methods for details).
Representative absorption images are shown in Fig.\,1c together with the extracted spin profiles  (green lines).  
The histograms in Fig.\,1c show the probability distribution of $F_x$ for all positions $y$ and  experimental realisations for the corresponding evolution time (see Extended Data Figure\,1 for all evolution times). 
Results are presented for characteristic stages associated to the initial condition (1), nonequilibrium instability regime (2), universal scaling regime (3) and departure from the non-thermal fixed point (4), as also indicated in Fig.\,1a.

\begin{figure*}
    \linespread{1}
	\centering
	\includegraphics[width = 2\columnwidth]{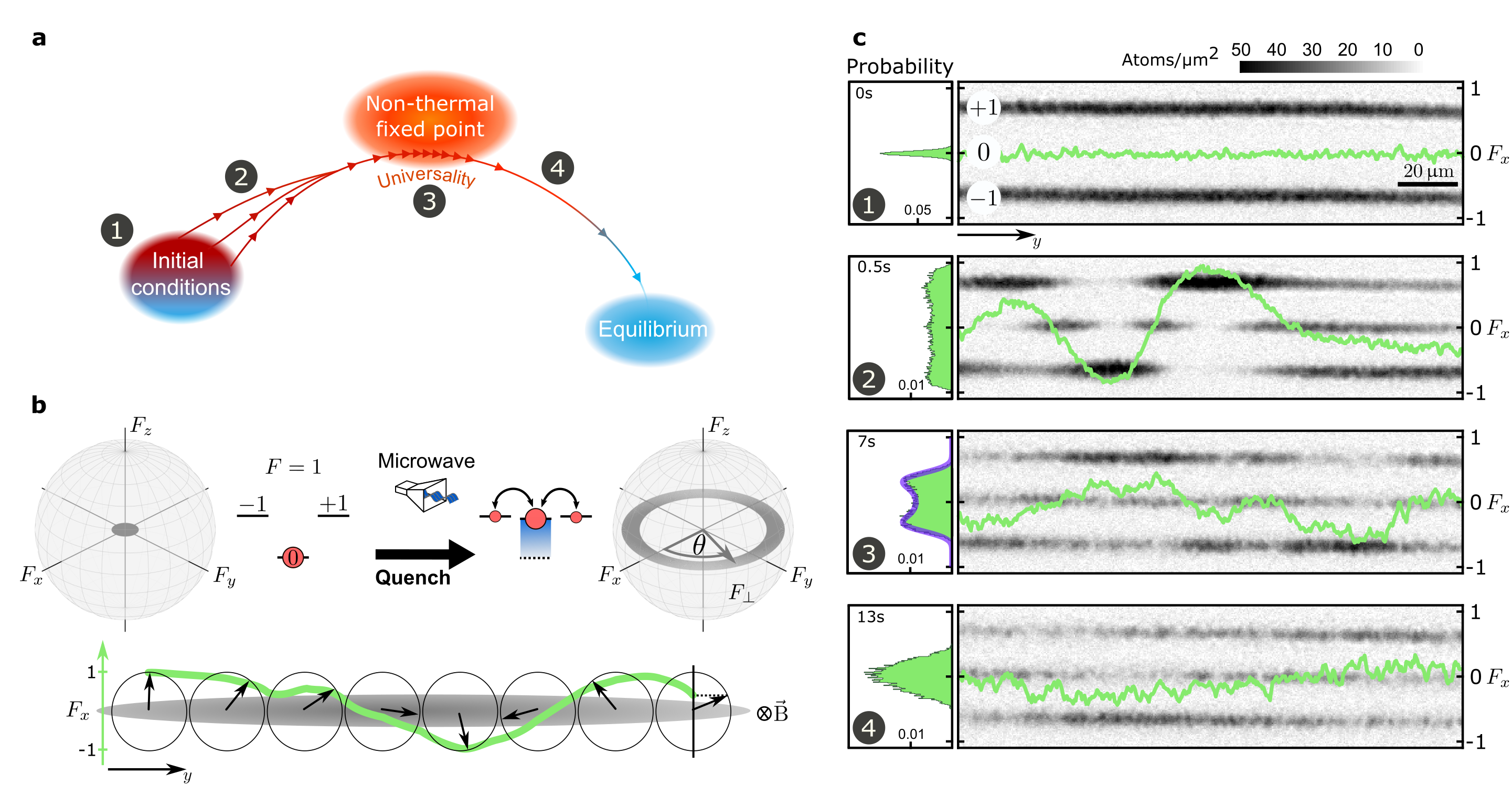}
	\caption{\textbf{Figure 1. Universal dynamics and experimental procedure}.
	 \textbf{a, } Starting from a class of far-from-equilibrium initial conditions, universal dynamical evolution indicates the emergence of a non-thermal fixed point.
	 Experimentally, we probe the system at different evolution times during the stages indicated by numbers 1 to 4.
\textbf{b, }A condensate is prepared in the $m_\text{F} = 0$ state of the $F = 1$ hyperfine manifold, i.e.~with a vanishing mean spin length (left spin sphere). 
         With microwave dressing (see Methods) we initiate spin-exchange dynamics which leads to a growth of spin orthogonal to the magnetic field $\vec{B}$ in the $F_{x}$--$F_{y}$-plane (right sphere). Subsequently, spatial structures of the spin orientation $\theta$ are found along the cloud.
\textbf{c, } Exemplary absorption images of the three hyperfine levels taken after a $\pi/2$ spin rotation and Stern-Gerlach separation  together with the inferred local spin $F_{x}(y)$ (green lines). Furthermore histograms for ${\sim160}$ experimental realisations are shown. 
        In the universal regime (see step 3 in panel a) we extract the spin length and its fluctuation by a fit to the double-peaked structure of the histogram, as indicated in the corresponding plot (see Methods).}
\end{figure*}

We find that during the time evolution the angular orientation $\theta$ of the transverse spin (see Fig.\,1b) becomes the relevant dynamical degree of freedom.
For short evolution times unstable longitudinal spin modes grow exponentially~\cite{Leslie2009}, well described by Bogoliubov theory, but non-linear evolution quickly takes over ($\gtrsim100\,$ms). 
This leads to a double-peaked structure of the histograms (see Fig.~1c) indicating that the spin has a mean length and a random orientation in the $F_{x}$--$F_{y}$-plane.
On the basis of this observation we extract the mean spin length $\langle |{F}_{\perp}(t)|\rangle$, where ${F}_{\perp}=F_{x}+iF_{y}$, and its fluctuations using a fit.
Building on that knowledge, we extract the local angle from the profiles as $\theta(y,t) = \arcsin(F_{x}(y,t)/\langle |{F}_{\perp}(t)|\rangle)$ (see Methods for details).

The time evolution of the fluctuations of the spin orientation is described in terms of correlation functions of the scalar field $\theta(y,t)$. 
The fluctuations are analysed by evaluating the two-point correlation function $C(y,y';t)=\langle\theta(y,t)\theta(y',t)\rangle$.
To distinguish the role of different length scales we consider a momentum-resolved picture of the dynamics.
Hence we evaluate the structure factor, which is the Fourier transform of $C(y,y';t)$ with respect to the relative coordinate $\bar{y} = y'-y$, averaged over y,
\begin{equation}
  f_{\theta}(k,t) = \iint\,\mathrm{d}{y}\,\mathrm{d}{\bar{y}}\,C({y}+{\bar{y}},{y};t)\,\exp( -i\,2\pi{k}{\bar{y}})\,.
\label{eq:StructureFactor}
\end{equation}
In general, the structure factor $f_\theta$ is a function of momentum $k$ which evolves in time $t$ in a way determined by the system parameters and initial conditions.
In Fig.\,2a, we plot $f_{\theta}({k},t)$ as a function of $k$ on a double-logarithmic scale for times between 4\,s and 9\,s. 
A characteristic shift of the structure factor towards smaller momenta as well as an increase of the low-momentum amplitude with time is observed.

\begin{figure*}
    \linespread{1}
	\centering
	\includegraphics[width = 2\columnwidth]{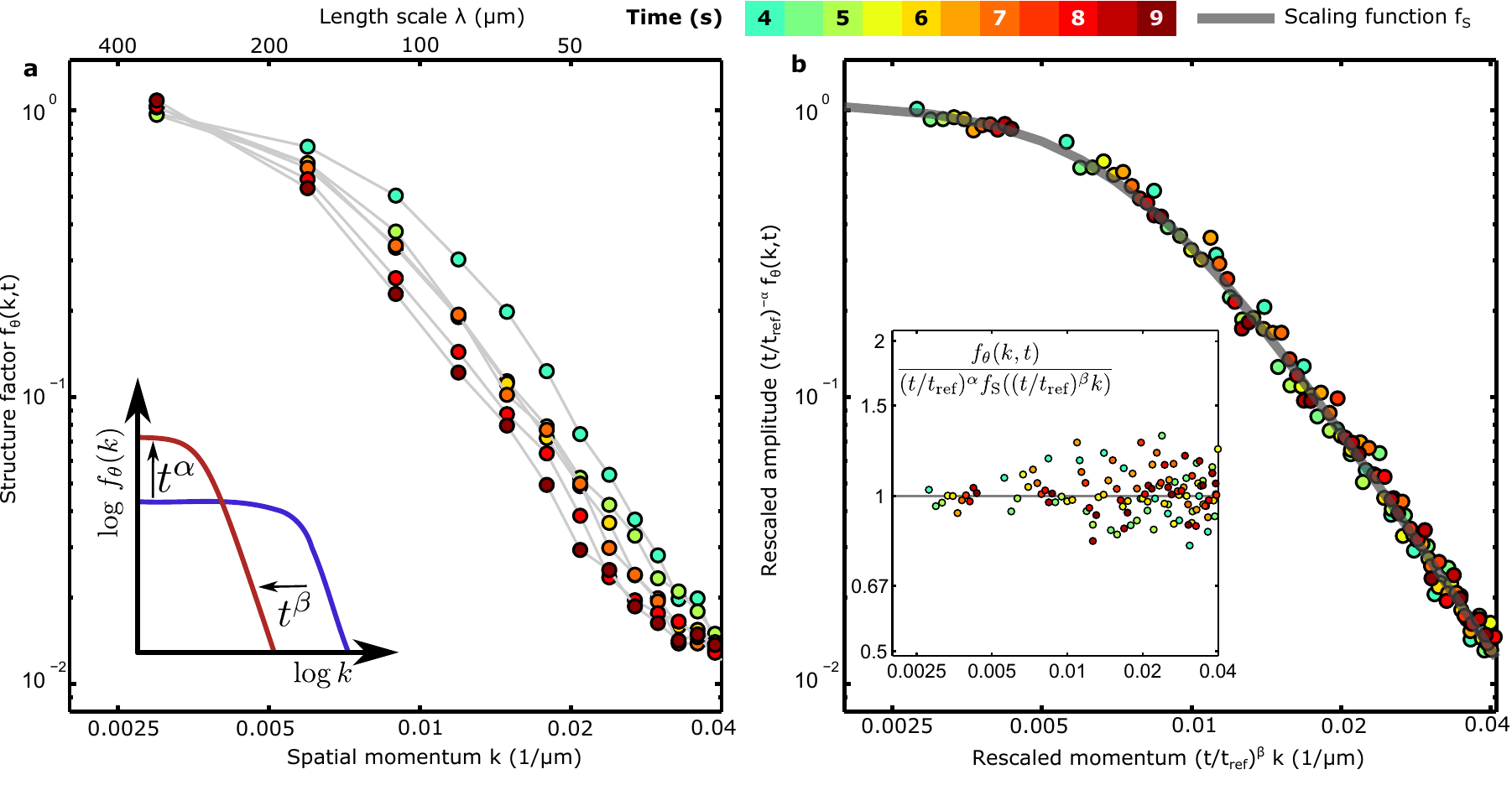}
	\caption{\textbf{Figure 2. Scaling in space and time at a non-thermal fixed point. a, } 
		Structure factor $f_\theta (k,t)$ as a function of the spatial momentum $k = 1/\lambda$ in the scaling regime between $4\,$s and $9\,$s. The color indicates the evolution time $t$. The statistical error is on the order of the size of the plot markers. In the infrared the structure factor shifts in time to smaller $k$ (bigger wavelengths) which is connected to transport of excitations towards lower momenta.  Characteristic for the non-thermal fixed point dynamics is the rescaling of the amplitude with universal exponent $\alpha$ and rescaling of the length scale with $\beta$ (see inset). \textbf{b, } By rescaling the data with $t_\text{ref} = 4.5\,$s,  $\alpha = 0.33$ and $\beta = 0.54$ the data collapses to a single curve. We parametrise the universal scaling function  with $f_S \propto 1/(1+(k/k_s)^\zeta)$. Using a fit (grey solid line) we find $\zeta \approx 2.6$ and $k_s \approx 1/133\,\upmu$m.  
The quality of the rescaling is revealed by the small and symmetric scatter of the rescaled data divided by the fit (see inset).}
\end{figure*}


In fact, instead of separately depending on $k$ and $t$ we find that in this regime the data sets collapse to a single curve if the rescaled distribution $t^{-\alpha} f_{\theta}$ is plotted as a function of the single variable $ t^\beta k$. 
This implies that the data satisfy the scaling form

\begin{equation}
  f_{\theta}({k},t) 
  =   t^\alpha f_{\text{S}}\! \left( t^\beta {k} \right)
\label{eq:ScalingHypothesis}
\end{equation}
with universal scaling exponents $\alpha$, $\beta$ and scaling function $f_S$.
Fig.\,2b shows this collapse, where the same data points as in Fig.\,2a are plotted with times normalised to the reference time $t_\text{ref} = 4.5\,$s. 
The ability to reduce the full nonequilibrium time evolution of the correlation function in the scaling regime to a time-independent, so-called fixed point distribution $f_S({k})$ and associated scaling exponents is a striking manifestation of universality. 

We find for the amplitude scaling exponent $\alpha = 0.33\pm0.08$ and for the momentum scaling exponent $\beta = 0.54\pm0.06$. 
The errors correspond to one standard deviation obtained from a resampling technique\,(see Methods).
However, the actual uncertainty for $\alpha$ is expected to be larger since the rescaling analysis is  much less constraining on $\alpha$ than on $\beta$.
We find that $f_{\theta}(k,t)$ develops a plateau at the lowest momenta and an approximate power-law fall-off above a characteristic length scale in the scaling regime. 
To parametrise the universal scaling function, we fit the rescaled data with a function of the form $f_S(k) \propto 1/[1+(k/k_s)^\zeta]$ \citep{Karl2016} and find $\zeta \approx 2.6$, with $k_s \approx 1/133\,\upmu \text{m} $ for our system. 
The value of $\zeta$ becomes constant after about 4\,s  (see Fig.\,3a).
Analysing $f_{\theta}(k=0,t)$ as shown in Fig.\,3b reveals that the occupation of $k=0$, which cannot be seen on the logarithmic scale employed in Fig.\,2, builds up in the scaling regime.
This growth is consistent with the power law $\sim t^\alpha$  with $\alpha$ obtained from the rescaling analysis as indicated by the solid green line.
After $9\,$s the system departs from the scaling behaviour.

\begin{figure*}
    \linespread{1}
	\centering
	\includegraphics[width = 1.5\columnwidth]{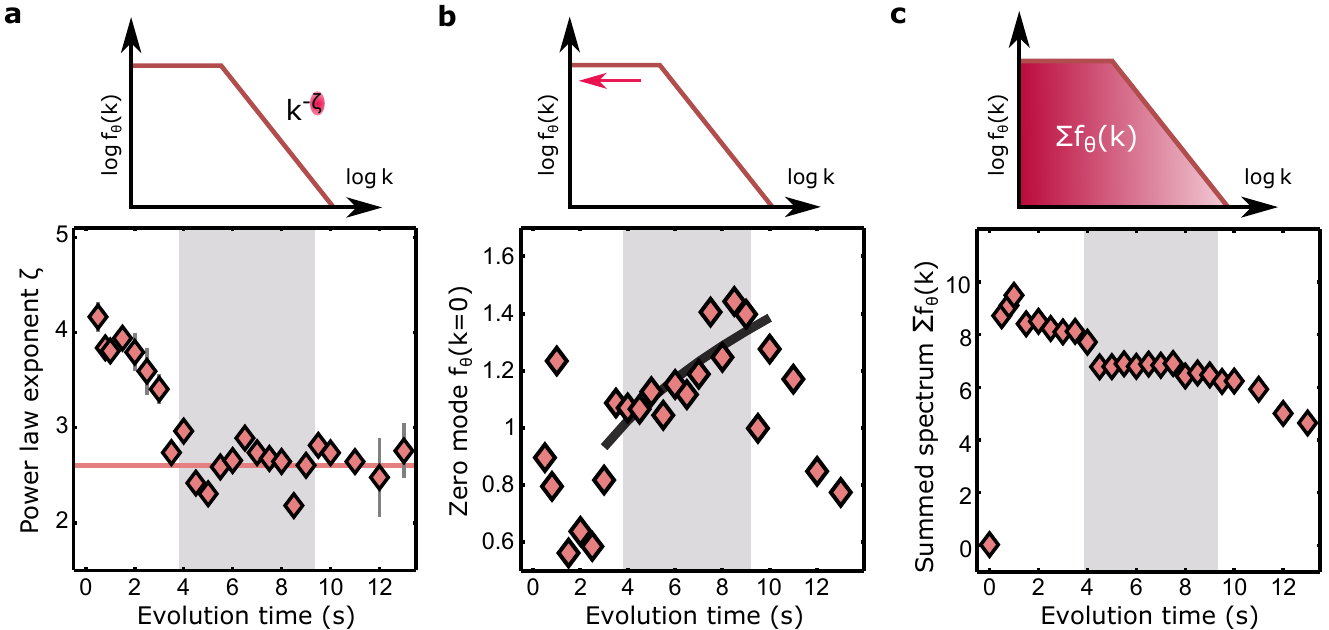}
	\caption{\textbf{Figure 3. Characterisation of the scaling regime.} 
\textbf{a, }For each evolution time (cf. Fig.\,2) we extract the power-law exponent $\zeta$ from a fit. After 4\,s it settles to $\approx 2.6$ (red solid line) revealing the build-up of the universal scaling function. 
The grey shaded region indicates the scaling regime. 
\textbf{b, } The transport to the infrared in the scaling regime is connected to a monotonic increase of the occupation of $k=0$. The solid line depicts the expected scaling $f_\theta(k=0,t) \propto {t}^\alpha$ with $\alpha=0.33$. 
After $9\,$s a rapid decay signals the departure from the scaling regime.  
 \textbf{c, } 
 The emergence of a conserved quantity is signalled by the sum over all $k$-modes of $f_\theta(k,t)$. 
After a fast initial growth this observable is approximately constant in the scaling regime and starts to decay after 9\,s.}
\end{figure*}


The nature of the observed scaling phenomenon is explained by the emergence of an approximately conserved quantity and its transport.
In terms of our dynamical degree of freedom $\theta(y,t)$ we identify $\int\mathrm{d}k\, \braket{|\theta(k,t)|^2} \equiv \int\mathrm{d}k\, f_{\theta}({k},t)$ as the conserved quantity.
In fact, Fig.\,3c shows that the sum over all modes $k$ for different evolution times -- after a fast initial rise due to the instability -- settles around a constant within the scaling regime (see also Extended Data Figure\,2).
According to the scaling (\ref{eq:ScalingHypothesis}), $\int\mathrm{d}k\, f_{\theta}({k},t)  = t^{\alpha-\beta}\int\mathrm{d}{k}\,f_S(k)\simeq const.\,$ corresponds to $\alpha \simeq \beta$ such that in our case only one independent dynamical scaling exponent remains.
A distinct feature is the transport of the conserved quantity directed towards the infrared corresponding to a positive sign of $\beta$.
Theoretically it is expected to find the scaling only for momenta smaller than some scale \citep{Orioli2015} (in our case $\sim0.04\,\mu \text{m}^{-1}$, see Extended Data Figure 3).
The transport towards the infrared is in contrast to the turbulent transport into the ultraviolet observed in direct cascades \citep{Navon2016}.

These experimental findings of scaling behaviour, implying universality, allow the comparison with predictions in a variety of models in the non-thermal universality class, which is defined by the scaling function $f_S$ and $\alpha = d\beta$ for given spatial dimension $d$.
$N$ interacting Bose gases with equal intra- and interspecies Gross-Pitaevskii couplings are described by an $O(N)$ symmetric model.
This is closely related to $O(N)$ symmetric scalar models \cite{ZinnJustin2004a}, such as the relativistic Higgs sector of the Standard Model with $N = 4$ for $d = 3$.
For these types of models, both,  Gross-Pitaevskii and relativistic, a universal value of $\beta \approx 0.5$ has been predicted and found to be insensitive to the spatial dimension for $d \geq 2$ \cite{Orioli2015}.
This describes the self-similar transport of excitations of the relative phases between the components to lower wave numbers.
The scaling function $f_S$ is known to depend on dimensionality \cite{Chantesana2018} and has not yet been theoretically estimated for $d = 1$.
Our setup is the first realisation of an effective $N = 3$ model for the transport of conserved quantities associated to non-thermal fixed points in a quasi one-dimensional situation.
Finding scaling behaviour in one dimension was not expected and sheds new light on the concept of universality classes far from equilibrium.

We emphasise that the non-thermal scaling phenomenon studied here involves no fine-tuning of parameters.
This is in contrast to equilibrium critical phenomena that require a careful adjustment of system variables such as the temperature to a critical value \cite{Hohenberg1977}.
To illustrate this insensitivity we employ the high level of control of the atomic spin system and prepare three qualitatively different initial conditions (for details see Methods).
The corresponding absorption images of single realisations are shown in Fig.\,4a along with the Fourier transform of the spatial correlation function of $F_x(y)$. 

\begin{figure}
    \linespread{1}
	\centering
	\includegraphics[width = 1\columnwidth]{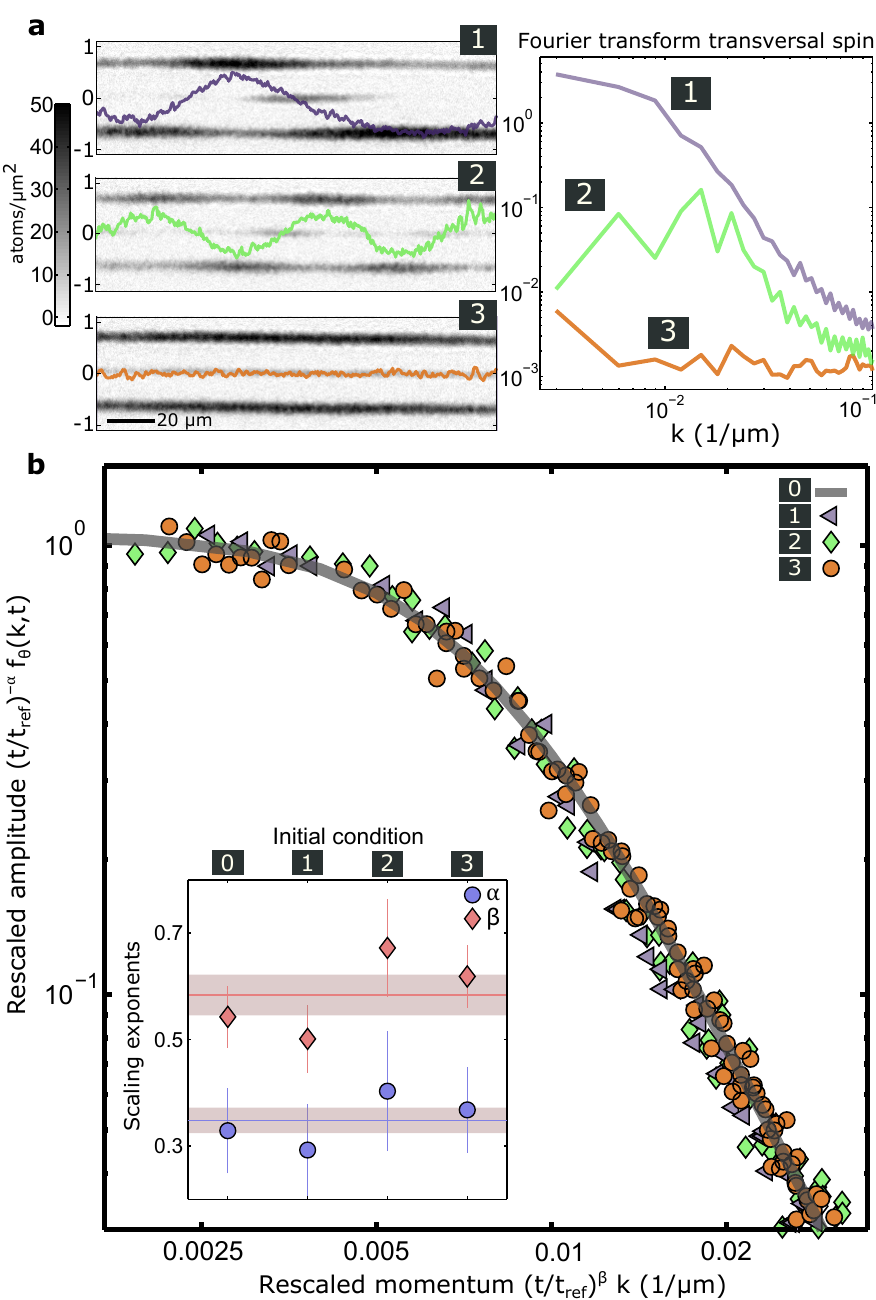}
	\caption{\textbf{Figure 4. Robustness of universal dynamics at a non-thermal fixed point. a, } Absorption images of all three $m_\text{F}$ components after spin rotation with the extracted transversal spin (solid lines) of three different initial conditions.
The preparations show different initial amplitudes in the Fourier transform of the transversal spin. \textbf{b, } All initial conditions lead to scaling dynamics. Data shown was obtained in a time window between $4\,$s and $9\,$s after preparation of the initial state. In the inset the scaling exponents of all four initial conditions, including the preparation in $m_\text{F} = 0$ (cf. Fig.\,1), are shown; the error bars are 1 s.d. obtained from a resampling method (see Methods). The mean values (red and blue solid lines) of $\alpha$ and $\beta$ are used to rescale the data.  We allow for overall scaling factors in $k$ and amplitude for each initial condition.}
\end{figure}

We find universal dynamics for all initial conditions with comparable inferred scaling exponents (see inset of Fig.\,4b).
We rescale the data with the same exponents obtained from the mean of all four measurements and take into account overall scaling factors and reference momentum scales.
This procedure leads to a collapse of all data manifesting the robustness of non-thermal fixed point scaling.

The demonstrated level of control and the accessible observables on our platform open the door to the discovery of further non-thermal universality classes. This represents a crucial step towards a comprehensive understanding of out-of-equilibrium dynamics with potential impact in various fields of science.

Similar phenomena have recently been observed by the Schmiedmayer group \cite{Erne2018} in Vienna in a single-component Bose gas where a scaling exponent $\beta \simeq 0.1$ was extracted.

\noindent\textbf{Acknowledgements}\\
We thank D. M. Stamper-Kurn, J. Schmiedmayer, A. Pi\~{n}eiro Orioli, M. Karl, J. M. Pawlowski and A. N. Mikheev for discussions.  \\
This work was supported by the Heidelberg Center for Quantum Dynamics, the European Commission FET-Proactive grant AQuS (Project No. 640800), the ERC Advanced Grant Horizon 2020 EntangleGen (Project-ID 694561) and the DFG Collaborative Research Center SFB1225 (ISOQUANT)\\
\noindent\textbf{Author contributions}\\
The experimental concept was developed in discussion among all authors.  M.P., P.K. and S.L. controlled the experimental apparatus. M.P., P.K., H.S., S.L. and M.K.O. discussed the measurement results and analysed the data. C.-M. S., J.B. and T.G. elaborated the theoretical framework. All authors contributed to the discussion of the results and the writing of the manuscript.

\noindent\textbf{Competing financial interests}\\
The authors declare no competing financial interests.

\noindent\textbf{Data availability}\\
The data presented in this paper are available from the corresponding author upon reasonable request.

\noindent\textbf{Author information}\\
Correspondence and requests should be addressed to M.P. (universaldynamics@matterwave.de).



\vspace*{0.5cm}
\begin{center}
  \textbf{Methods}
\end{center}
\vspace*{0.5cm}

\subsection{Microscopic parameters}
The dynamics of the spinor Bose gas is described by the Hamiltonian
\begin{align}
\hat{H} 
=&  \hat{H}_0 + \int  \! \mathrm{d}V \left[ \, : \frac{c_0}{2} \hat{n}^2 +\, \frac{c_1}{2} \left(  \hat{F_x}^2+ \hat{F_y}^2+ \hat{F_z}^2 \right): + q \left( \hat{n}_{+1} + \hat{n}_{-1}  \right) + p\hat{F}_z \right]
\end{align}

where $\psi_{{m}}^{\dag}$ is the bosonic field creation operator of the magnetic substate $m\in \{ 0,\pm1 \}$ and  $\hat{n}_m=\hat{\psi}_{m}^{\dag}\hat{\psi}_{{m}}$ and $::$ denotes normal ordering. 
$\hat{H}_0$ contains the spin independent kinetic energy and trapping potential.
The spin operators are given by: $\hat{F}_x =  \left[ \hat{\psi}_0^\dag \left( \hat{\psi}_{+1} + \hat{\psi}_{-1} \right) + \text{h.c.} \right] / \sqrt{2}$ and $\hat{F}_y =  \left[ i \hat{\psi}_0^\dag \left( \hat{\psi}_{+1} - \hat{\psi}_{-1}\right) + \text{h.c.} \right] / \sqrt{2}$ and $\hat{F}_z = \hat{n}_{+1}-\hat{n}_{-1}$.
The parameter $p$ describes the linear Zeeman shift in a magnetic field.
For the hyperfine spin $F = 1$ of $^{87}$Rb the spin interaction is ferromagnetic, i.e. $c_1<0$.

For the experimental control parameter $q>2n|c_1|$, with $n$ being the total density, the mean-field ground state is the polar state, which corresponds to all atoms occupying the $m_\text{F} = 0$ state.
In the range $0<q<2n|c_1|$ a spin with non-vanishing length in the $x$-$y$-plane is energetically favoured (easy-plane ferromagnet) \citep{Kawagushi2012}.
This is the parameter regime employed in the experiment.

\subsection{Experimental system}
We prepare a BEC of $\sim 70,000$ atoms in the state $(F,\,m_\text{F}) = (1,0)$ in an optical dipole trap of $1030\,$nm light with trapping frequencies $(\omega_\parallel,\omega_\perp) \approx 2\pi\times(2.2,250)\,$Hz.

The control parameter $q$ is given by $q = q_B - q_\text{MW}$, where  $q_B \approx 2\pi \times 56\,$Hz  is the second-order Zeeman splitting at a magnetic field of $B \approx 0.884\,$G and $q_\text{MW} = \Omega^2/4\delta$ is the energy shift due to the microwave dressing.
For dressing \cite{Gerbier2006} we use a power-stabilised microwave generator with resonant Rabi frequency $\Omega \approx  2\pi \times 5.3\,$kHz and $\delta \approx  2\pi \times 137\,$kHz blue detuned with respect to the $(1,0)\leftrightarrow (2,0)$ transition. 
For the spin dynamics we adjust $\Omega$ and $\delta$ such that $q\approx n|c_1|$  (with $nc_1 \approx  -2\pi \times 2\,$Hz).
In order to monitor the long-term stability of $q$ we do a reference measurement every $4\,$h (corresponding to $\sim250$ experimental realisations).
For this we observe spin dynamics for a fixed evolution time of 4\,s as a function of the control parameter $q$ (changing the detuning $\delta$).
Analysing the integrated side mode population we infer that the drifts of $q$ are well below $0.5\,$Hz. 

\subsection{Preparation of different initial conditions}

We prepare three initial conditions (ICs, see Fig.\,4) which differ from the polar state.
For IC\,1  the control parameter is first set to $q\approx n|c_1|+1\,$Hz. 
After 500\,ms of spin dynamics at this value we quench to the final value $q\approx n|c_1|$. 
For the preparation of IC\,2 we apply a resonant $\pi/5$ radiofrequency (rf) pulse to populate the $(1,\pm1)$ states.
After a hold time of 100\,ms at a magnetic field gradient of $\approx 0.2\,\upmu \text{G}/\upmu \text{m}$ in the longitudinal trap direction we apply a second $\pi/5$ rf-pulse.
The combination of $q$ and an inhomogeneous $p$ during the hold time leads to a  spatially modulated transversal spin on a length scale of $\lambda \approx 80\,\upmu \text{m}$.
For IC\,3 we populate homogeneously the $(1,\pm1)$ states with a short rf-pulse such that $(n_{+1}+n_{-1})/n\approx0.1$.

\subsection{Spin read-out}
The spin dynamics is initiated by quenching the control parameter.
After a fixed evolution time $t$ we apply a short magnetic field gradient pulse (Stern-Gerlach) in $z$-direction and switch off the waveguide potential. 
Following a short time of flight ($\sim 1\,$ms) we perform high intensity absorption imaging  with a resonant light pulse of $15 \,\upmu$s duration.
The resolution of the imaging system is $\approx 1.2\,\upmu$m corresponding to three pixels on the CCD camera \citep{Muessel2013}; we accordingly bin the spin profiles by three pixels.
As our Stern-Gerlach analysis is oriented in $z$-direction, for the read-out of the spin in the $x$--$y$-plane we apply, prior to the magnetic field gradient, an rf-pulse resonant with the transitions $(1,0) \leftrightarrow (1,\pm1)$. 

The rf-pulse can be modelled as a spin rotation described by the Hamiltonian $\hat{H}_\text{rf} = \Omega_\text{rf} \hat{F}_y$ with resonant Rabi frequency $\Omega_\text{rf} \approx 2\pi \times 17.5\,$kHz.
Applying a $\pi/2 $-pulse of duration $ 14.3\,\upmu$s the observable $\hat{F}_x$ is mapped to the measurable density difference ${n}_{+1} - {n}_{-1}$.

\subsection{Inferring the spin orientation}
The double-peaked spin distributions in the scaling regime (see Extended Data Figure 1) resemble a distribution of a transversal spin with random orientation. 
To extract the corresponding ensemble average length $\braket{|F_\perp|}$ of the transversal spin and its fluctuation $\sigma$ we fit a probability density of the form $p(F_x) \propto 1/\sqrt{1-(F_x/\braket{|F_\perp|})^2}$ convolved with a Gaussian distribution with rms $\sigma$. 

Under the assumption of a homogeneous spin length  the spatial profile of the angular orientation is given by $\theta(y) = \arcsin(F_x(y)/\braket{|F_\perp|})$. 
If the maximal amplitude is larger than $\braket{|F_\perp|}-\sigma$ we use the maximal amplitude of the single realisation instead of $\braket{|F_\perp|}$.

\subsection{Extraction of scaling exponents}
After rescaling the results of the discrete Fourier transform according to eq. \eqref{eq:ScalingHypothesis} we interpolate with cubic splines to obtain a common $k$-grid for all evolution times.
We vary the scaling exponents $\alpha$ and $\beta$ to minimise the sum of the squared relative differences of all structure factors $f_\theta$.
To estimate the statistical error on the exponents we employ a jackknife resampling analysis \citep{Miller1974}.

\makeatletter
\apptocmd{\thebibliography}{\global\c@NAT@ctr 31\relax}{}{}
\makeatother

\vspace*{0.5cm}
\begin{center}
  \textbf{Extended Data Figures}
\end{center}
\setcounter{figure}{0}
\renewcommand{\figurename}{\textbf{Extended Data Figure}}

\vspace*{0.5cm}
\newpage
\begin{figure*}
    \linespread{1}
	\centering
	\includegraphics[width = 2\columnwidth]{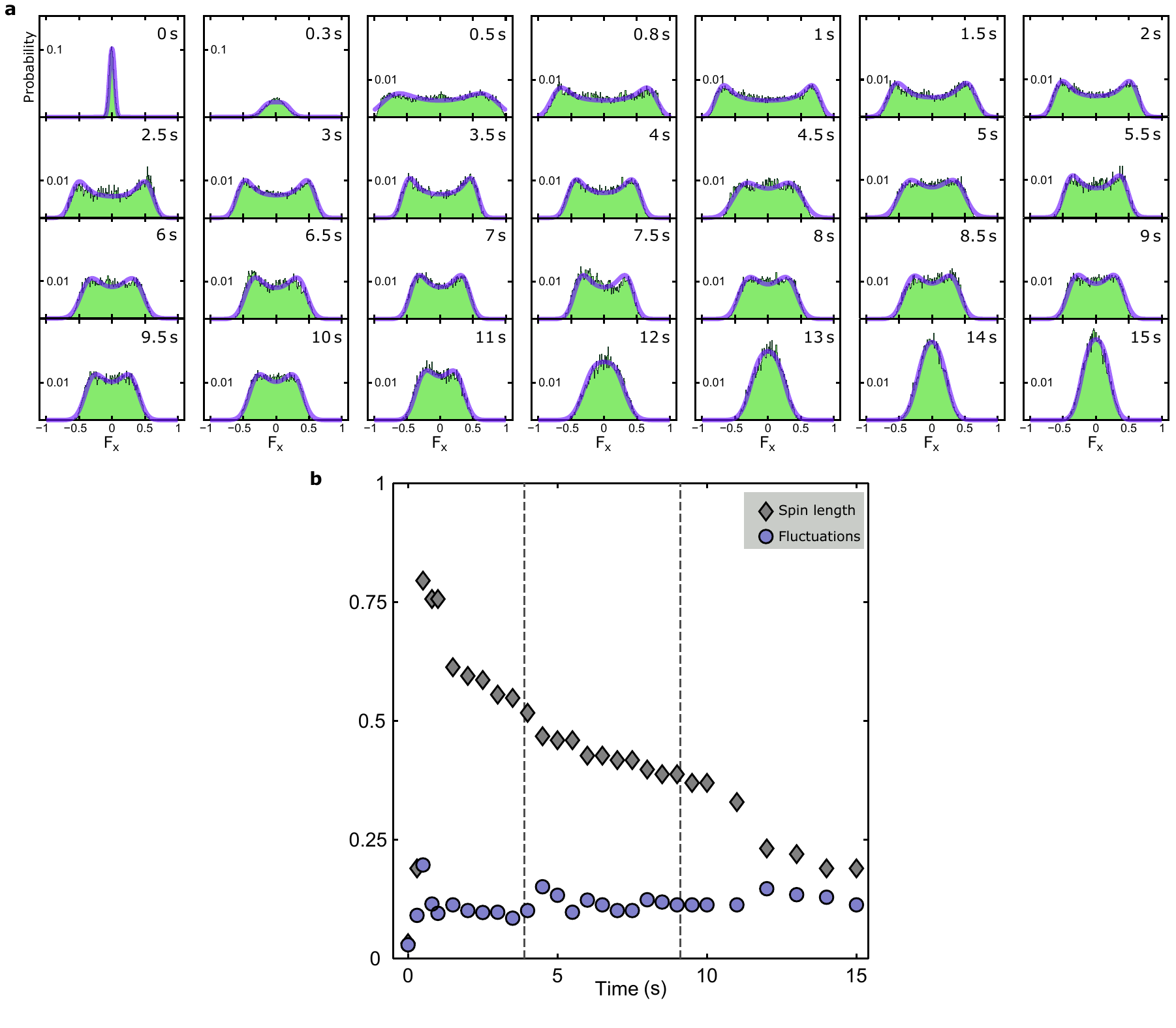}
	\caption{\textbf{Extended Data Figure 1. Spin distributions for all evolution times.} \textbf{a)} The panels show the distributions of the transversal spin $F_x$ measured at different evolution times as indicated. 
	Initially, we find a narrow Gaussian distribution corresponding to the prepared coherent spin state. The excitations developing in the transversal spin lead to a double-peaked distribution within the interval of $2\,$s to $10\,$s. 
	For long evolution times, $t>12\,$s, the distribution resembles a Gaussian, which is much broader than the initial distribution. \textbf{b)} The spin length and its rms fluctuation as a function of evolution time are extracted by a fit (see Methods). We find a slow decay of the spin length and nearly constant rms fluctuations in the scaling regime.}
	\end{figure*}

\newpage
\begin{figure}
    \linespread{1}
	\centering
	\includegraphics[width = 1\columnwidth]{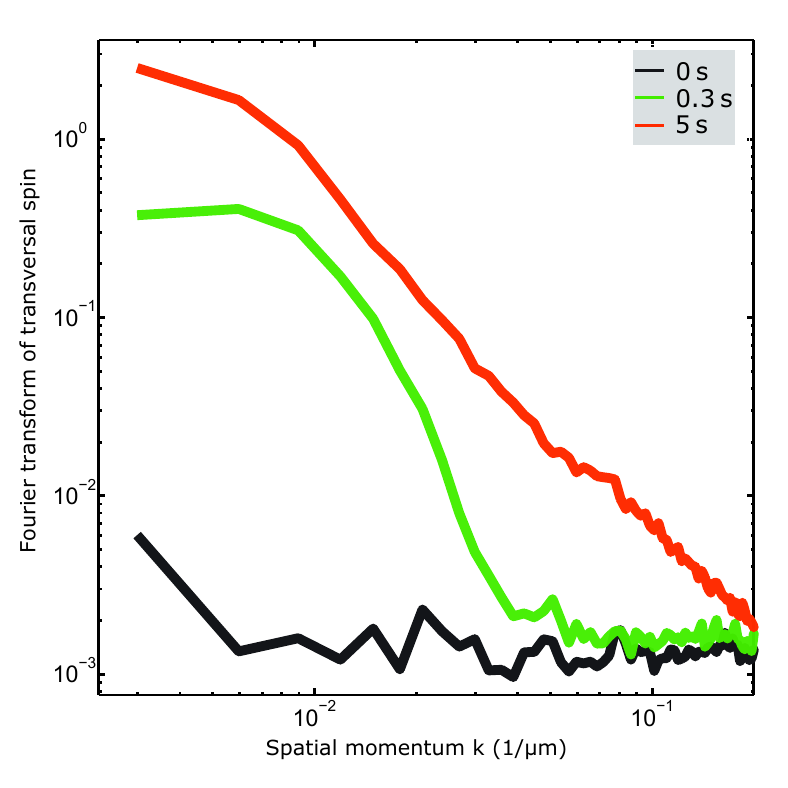}
	\caption{\textbf{Extended Data Figure 2. Build-up of transversal spin in momentum space.} Since the angular orientation $\theta$ cannot be extracted reliably for short evolution times, we choose to show the  Fourier transform of the transversal spin for regimes 1-3 (cf. Fig.~1). 
	The initial condition, all atoms prepared in $m_\text{F} = 0$, is characterised by a flat distribution. There is a fast build-up of long-wavelength spin excitations by more than two orders of magnitude within the first second. This process is followed by a redistribution of momenta leading to the scaling form for times longer than $4\,$s.}
	\end{figure}


\newpage
\begin{figure*}
    \linespread{1}
	\centering
	\includegraphics[width = 2\columnwidth]{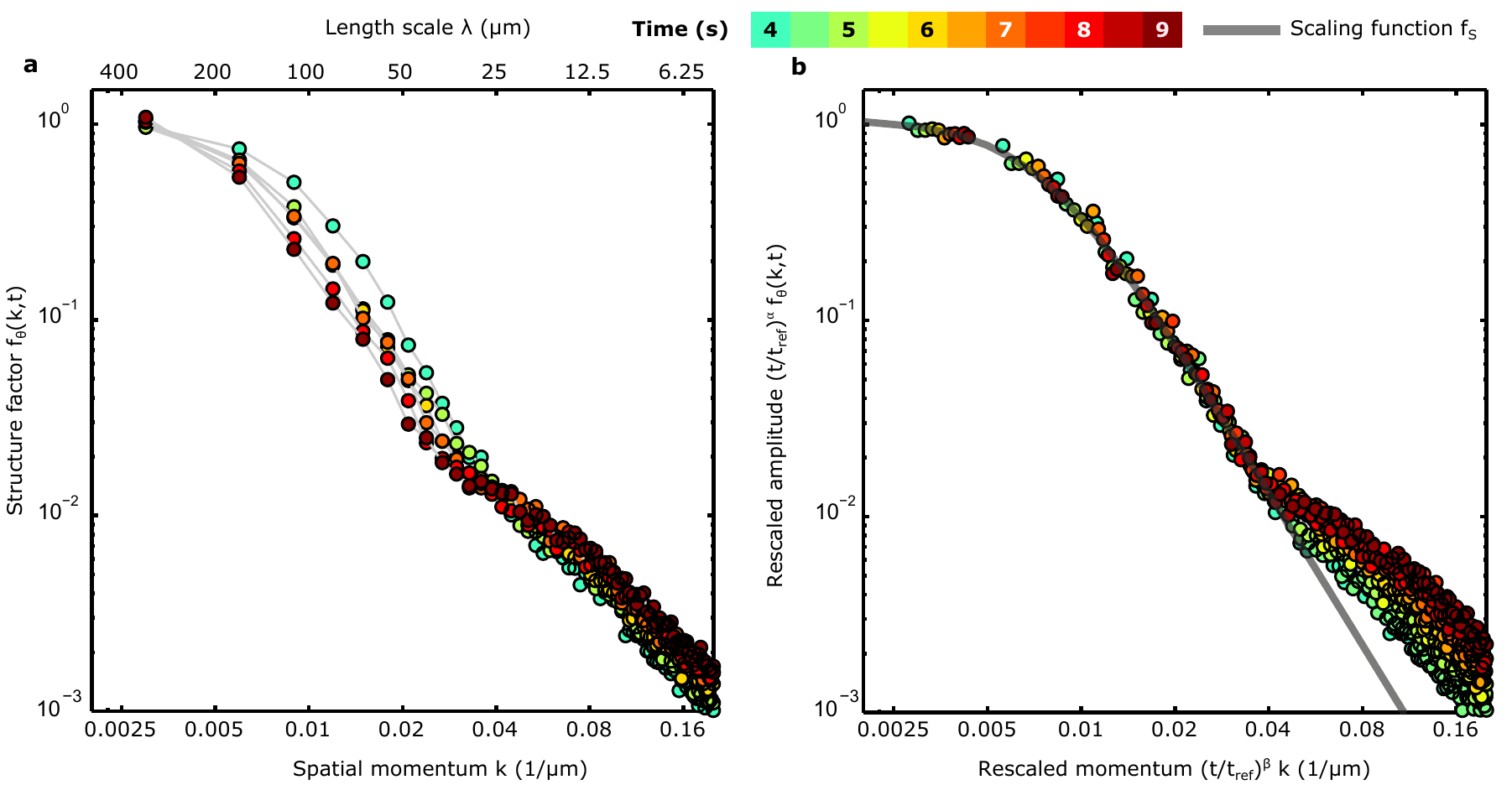}
	\caption{\textbf{Extended Data Figure 3. Scaling of structure factor for all experimentally accessible length scales.} Same data as shown in Fig.~2. The rescaling does not apply for large momenta $k>0.04\,\mu\text{m}^{-1}$.}
	\end{figure*}


\clearpage



\end{document}